%% file: iclr2022_workshop.tex
\documentclass{article}
\usepackage{iclr2022_workshop,times}

\input{math_commands.tex}

\usepackage{hyperref}
\usepackage{url}
\usepackage{xcolor}
\usepackage{enumitem}

\title{
\centering
ICLR 2022 Challenge for 

Computational Geometry \& Topology:

Design and Results}

\author{Adele Myers* \\
UC Santa Barbara, USA\\
\texttt{adele@ucsb.edu} \\
\And
Saiteja Utpala* \\
UC Santa Barbara, USA\\
\And
Shubham Talbar* \\
UC Santa Barbara, USA\\
\And
Sophia Sanborn* \\
UC Berkeley, USA\\
\And
Christian Shewmake* \\
UC Berkeley, USA\\
\And
Claire Donnat* \\
The University of Chicago, USA\\
\And
Johan Mathe* \\
Atmo, USA \\
\And
Umberto Lupo* \\
EPFL, Switzerland \\
\And
Rishi Sonthalia** \\
UCLA, USA \\
\And
Xinyue Cui** \\
UCLA, USA \\
\And
Tom Szwagier** \\
ENS Paris-Saclay, France\\
\And
Arthur Pignet** \\
ENS Paris-Saclay, France \\
\And
Andri Bergsson** \\
PayAnalytics 
Iceland \\
\And
Søren Hauberg** \\
Technical University of Denmark, Denmark \\
\And
Dmitriy Nielsen** \\
University of Copenhagen, Denmark\\
\And
Stefan Sommer** \\
University of Copenhagen, Denmark\\
\And
David Klindt** \\
NTNU, Norway \\
\And
Erik Hermansen** \\
NTNU, Norway \\
\And
Melvin Vaupel** \\
NTNU, Norway \\
\And
Benjamin Dunn** \\
NTNU, Norway \\
\And
Jeffrey Xiong ** \\
Columbia University, USA \\
\And
Noga Aharony ** \\
Columbia University, USA \\
\And 
Itsik Pe'er ** \\
Columbia University, USA \\
\And
Felix Ambellan** \\
Freie Universität Berlin, Germany \\
\And
Martin Hanik** \\
Freie Universität Berlin, Germany \\
\And
Esfandiar Nava-Yazdani** \\
Zuse Institute Berlin, Germany \\
\And
Christoph von Tycowicz** \\
Freie Universität Berlin, Germany \\
\And
Nina Miolane* \\
UC Santa Barbara, USA\\
\texttt{ninamiolane@ucsb.edu} \\
\AND
   \\
\vspace{-3mm}
\** : Organizers and external jury; \**\** : Participants
}

\iclrfinalcopy 
\begin{document}

\maketitle

\begin{abstract}
    This paper presents the computational challenge on differential geometry and topology that was hosted within the ICLR 2022 workshop ``Geometric and Topological Representation Learning". The competition asked participants to provide implementations of machine learning algorithms on manifolds that would respect the API of the open-source software Geomstats (manifold part) and Scikit-Learn (machine learning part) or PyTorch. The challenge attracted seven teams in its two month duration. This paper describes the design of the challenge and summarizes its main findings.
    \textbf{Code:} \url{https://github.com/geomstats/challenge-iclr-2022}. \textbf{DOI:} 10.5281/zenodo.6554616.
\end{abstract}

\section{Introduction}

Traditional statistics and machine learning have been developed for data that lie on Euclidean vector spaces. \textit{Geometric statistics} and \textit{geometric machine learning} extend traditional methods to data that lie on nonlinear spaces such as manifolds. Geometric methods use the geometry of the data space as an inductive bias to guide modeling and analysis. This approach has shown great promise in many application domains, ranging from biomedical imaging to palaeontology. However, to date, the adoption of geometric methods has been  limited by the scarce availability of open-source and unit-tested implementations of published algorithms. The challenge described in this white paper aimed to fill that gap.

The purpose of this challenge was to foster reproducible research in geometric machine learning, by crowd-sourcing the open-source implementation of learning algorithms on manifolds, as motivated in the ICLR 2021 challenge's white paper \citep{miolane2021iclr}. Participants were asked to contribute code for a published or unpublished algorithm, following Scikit-Learn/Geomstats' or PyTorch's APIs and computational primitives, benchmark it, and demonstrate its use in real-world scenarios. 

This white paper is organized as follows. Section~\ref{sec:setup} describes the setup of the challenge, including its guidelines and evaluation criteria. Section~\ref{sec:results} summarizes the submissions to the challenge, and Section~\ref{sec:features} describes the principle take-aways of the challenge for the development of geometric learning via the software Geomstats. Section~\ref{sec:rank} provides the final ranking of the submissions.

\section{Setup of the challenge}\label{sec:setup}

The challenge was held in conjunction with the workshop ``Geometric and Topological Representation Learning" of the International Conference on Learning Representations (ICLR) 2022 \footnote{Workshop ``Geometric and Topological Representation Learning": \url{https://gt-rl.github.io/}}. Participants were asked to contribute code for a geometric machine learning algorithm, following Scikit-Learn/Geomstats' or PyTorch's APIs and computational primitives, benchmark it, and demonstrate its use in real-world scenarios.

\paragraph{Guidelines} Each submission was required to take the form of a Jupyter Notebook. The participants were asked to submit their Jupyter Notebook via Pull Requests (PR) to the GitHub repository of the challenge\footnote{Challenge's repository: \url{https://github.com/geomstats/challenge-iclr-2022}}. This challenge requested Jupyter Notebook submissions because they offer authors a natural way to communicate results that are inherently reproducible, as anyone with access to the notebook may run the notebook to attain the same results.

Teams were accepted and there was no restriction on the number of team members. The principal developers of Geomstats (i.e. the co-authors of Geomstats published papers) were not allowed to participate. 

The participants were asked to include the following sections in their submission: 
\begin{itemize}[leftmargin=*]
    \item Introduction: Explain and motivate the choice of learning algorithm.
    \item Related work and implementations: Contrast the chosen learning algorithms with other algorithms, and describe existing implementations (if any).
    \item Implementation of the learning algorithm.
    \item Tests on synthetic datasets and benchmark.
    \item Application to real-world datasets.
\end{itemize}

\paragraph{Evaluation criteria: fostering creativity} The evaluation criteria were:
\begin{enumerate}[leftmargin=*]
    \item How ``interesting"/``important"/``useful" is the learning algorithm? Note that this is a subjective evaluation criterion, where the reviewers evaluated what the implementation of this algorithm brings to the community (regardless of the quality of the code).
    \item How readable/clean is the implementation? How well does the submission respect Scikit-Learn/Geomstats/PyTorch's APIs? If applicable: does it run across backends?
    \item Is the submission well-written? Do the docstrings help understand the methods?
    \item How informative are the tests on synthetic data sets, the benchmarks, and the real-world application?
\end{enumerate}

Note that these criteria were not aimed to reward new learning algorithms, nor learning algorithms that outperform the state-of-the-art. Instead, these criteria were designed to reward clean code and exhaustive tests that will foster reproducible research in geometric learning.

\paragraph{Software engineering practices} The participants were also encouraged to use software engineering best practices. Their code should be compatible with Python 3.8 and make an effort to respect the Python style guide PEP8. The \texttt{Jupyter notebooks} were automatically tested when a Pull Request was submitted and the tests were required to pass. If a dataset was used, the dataset had to be public and referenced. Participants could raise GitHub issues and/or request help or guidance at any time through the \texttt{Geomstats} slack workspace. Help and guidance was be provided modulo availability of the maintainers.

\paragraph{Comparison with the ICLR 2021 challenge} The ICLR 2022 challenge described here has key differences with the challenge organized in 2021. The ICLR 2021 challenge was purely open-ended; by contrast, this ICLR 2022 challenge was more guided, as to foster computational developments into a restricted area of computational geometry: machine learning on manifolds.

\section{Submissions to the Challenge}\label{sec:results}

Seven teams participated in the challenge by submitting code before the deadline:
\begin{itemize}
    \item NeuroSEED in Small Open Reading Frame Proteins (NeuroSEED),
    \item Wrapped Gaussian Process Regression on Riemannian Manifolds (WGPR),
    \item Riemannian Stochastic Neighbor Embedding (Rie-SNE),
    \item Topological Ensemble Detection with Differentiable Yoking (TEDDY),
    \item Sampler for Brownian Motion on Manifolds (Brownian Motion Sampler),
    \item Hyperbolic Embedding via Tree Learning (Tree Embedding),
    \item Sasaki Metric and Applications in Geodesic Analysis (Sasaki).
\end{itemize}

This section provides a summary of the submissions. In these summaries, we group the submissions into three categories:
\begin{itemize}
    \item General Methods: these submissions implement tools that can be applied to any Riemannian Manifold (WGPR, Rie-SNE, Brownian Motion Sampler).
    \item Metric-Specific: these submissions focus on implementing a specific metric (Tree Embedding, Sasaki Metric).
    \item Application Driven: for these submissions, the geometry is present in the data, and a geometric machine learning method has been designed for a particular application (TEDDY, NeuroSEED).
\end{itemize}

\subsection{General Methods}

These submissions implement general tools which could theoretically be applied to any manifold.

\paragraph{Riemannian Stochastic Neighbor Embedding (Rie-SNE)} Stochastic neighbor embedding (SNE) \citep{sne, tsne} aims to represent high-dimensional data clusters in lower dimensional space to make clustering visualization easier. The vast majority of dimensionality reduction techniques assume that data resides on a Euclidean domain, which presents problems when data lie in more complex spaces. This submission developed an extension of SNE that generalizes SNE to Riemannian manifolds \citep{riesne}. Rie-SNE reduces the dimensionality of a set of data by ``projecting" the high dimensional data onto a lower dimensional manifold, where data is easier to visualize. Rie-SNE does this by using a Riemannian probability distribution rather than a Gaussian one when computing high-dimensional similarities between data points. The team tested their submission on synthetic data (MNIST data points mapped to a 784 dimensionsional sphere) and compared their manifold clustering visualization result to results from a traditional data visualization algorithm: tangent PCA \citep{Fletcher2004}. They found that the clustering structure in every set of data is significantly more evident in Rie-SNE than it is in tangent PCA. Tangent PCA distorted structure while Rie-SNE did not. 

\paragraph{Wrapped Gaussian Process Regression on Riemannian Manifolds (WGPR)} This submission implements a method for nonlinear regression on Riemannian manifolds. Although regression for datasets lying on Euclidean spaces is a well-established field, not many methods exist for data on Riemannian manifolds; they are either too simple like Geodesic Regression \citep{thomas_fletcher_geodesic_2013}, or they lack interpretability. Gaussian Process Regression (GPR) \citep{gortler2019a} is a well-known nonlinear regression method which incorporates prior knowledge about the data distribution in Euclidean spaces. \citep{mallasto_wrapped_2018} generalize it to data lying on Riemannian manifolds using manifold-preserving tools, such as the ones implemented in \texttt{Geomstats}. This submission proposes an implementation of the latter's ``Wrapped Gaussian Process Regression" in \texttt{Geomstats} and shows that it has many advantages over related Euclidean and geodesic models on synthetic examples (toy data on a sphere) as well as real life ones (diffusion Tensor Imaging in the corpus callosum).

\paragraph{Sampler for Brownian Motion on Manifolds} Brownian motion, also known as the ``random walk", is common in physical systems. The scope of this submission is to implement a sampler for Brownian motion with non-trivial covariance on manifolds in \texttt{Geomstats}. In other words, the submission designs a random walk such that the possible (random) positions at an arbitrary time $t$ are distributed according to a distribution of choice (for example, Normal distribution). They achieve this via stochastic development of Euclidean Brownian motion using the bundle of linear frames (the frame bundle). The project demonstrates implementation on a two-dimensional sphere, but their method should work on general manifolds.

\paragraph{Geomstats implementation} WGPR has been implemented into \texttt{Geomstats} as a tool for manifold regression. Rie-SNE provides a suggestion for visualizing high-dimensional data clustering. Currently, Geomstats does not utilize stochastic development on any manifolds. The Brownian motion submission proposes one way to do this.

\subsection{Metric-Specific}

Both the Tree Embedding and the Sasaki Metric submissions proposed implementations of a specific metric. Both submissions can be applied to any Riemannian manifold.

\paragraph{The Tree Embedding} submission aims to provide a tool for visualizing tree-structured data in two dimensions. 
The Tree Embedding team utilized \texttt{TreeRep} \citep{NEURIPS2020_093f65e0} 
(a recent tree learning algorithm) and Sakar's algorithm \citep{MR2928298}
to embed data in tree structures and then reduce the dimensionality of the data set for ease of visualization. This submission will work best on data sets that lie on manifolds equipped with a metric that is close to a tree metric, but this submission is equipped to handle data with any structure. \texttt{TreeRep} is an algorithm which can either (i) take a tree metric and construct a tree structure or (ii) take a non-tree metric and construct a tree structure that best approximates a tree metric for that data. Their program learns a tree from a distance matrix, which represents data in a metric space. Then, their program embeds (weighted) trees in a 2D hyperbolic space (\texttt{PoincareDisk} from \texttt{Geomstats}) using Sakar's algorithm. The final result helps visualize data as a two-dimensional tree structure. They tested their pipeline on synthetic data generated by taking random points in a high-dimensional hyperbolic manifold. They also tested their pipeline on the Karate Dataset \citep{karateclub}
and the CS Ph.D. dataset \citep{denooy_mrvar_batagelj_2018}.
They showed that the popular ``optimization-based methods" take much longer and do not necessarily produce better results.

\paragraph{The Sasaki} submission implemented and tested a new \texttt{SasakiMetric} class, which will be implemented as a subclass of the \texttt{RiemannianMetric} class in \texttt{Geomstats}. The \texttt{SasakiMetric} subclass will be a valuable tool in \texttt{Geomstats} because the Sasaki metric is suitable for comparing geodesics on manifolds. It is also a natural choice of Riemannian structure for operating on tangent bundles of manifolds. Because the Sasaki metric is well-suited for comparing distances between geodesics on a manifold, it is extremely useful for comparing longitudinal data. As proof, the group tested their submission by calculating the mean trend for longitudinal rat scull data, from Vilmann's rat data set \citep{Bookstein1992}. In the study, eight landmark measurements were made on 18 rat sculls at various stages in the rat's life. Thus one ``trajectory" in this case corresponds to a sequence of landmark measurements performed on one rat throughout its lifetime. Removing scaling and rotations, this team represented the data as points in the Kendall's shape space. They then used geodesic regression to fit a geodesic to each trajectory, and geodesics were canonically identified with points in the tangent bundle. They used computational primitives of \texttt{Geomstats} to calculate the mean trajectory (with respect to the Sasaki metric on the tangent bundle of Kendall’s shape space) across all rats for each landmark, and they showed that this mean was reasonable.

\paragraph{Geomstats implementation} The Tree Embedding submission presents a 2D visualization method, which is akin to manifold embedding or Riemannian manifold learning within a metric space. In addition to its tree representation capabilities, this submission could be valuable for future \texttt{Geomstats} implementation by providing tools for generating a 2D hyperbolic space from data in a (tree-like) metric space. The \texttt{SasakiMetric} is currently being implemented in \texttt{Geomstats} as a subclass of \texttt{RiemannianMetric}. It will provide a new geometry that existing \texttt{Geomstats} methods can be run on, which will open additional doors to new machine learning opportunities.

\subsection{Application-Driven}

Application-driven submissions aim to implement a geometric learning algorithm for a particular type of data with a specific geometric structure. Both the TEDDY and NeuroSEED submissions fit in this category. 

\paragraph{TEDDY}provides a first proof-of-concept for the problem of clustering based on topological signatures in a dataset.
They built an unsupervised clustering algorithm that can separate neural ensembles tuned to distinct latent variable manifolds, including: $\mathbb{T}^2, \mathbb{S}^1, SO(3)$, and $\mathbb{R}^2$. 
In other words, TEDDY uses topological data analysis (TDA) to identify clusters in the dataset through their topological signatures. 
Note that this algorithm does not learn the manifold itself, but it does learn the topology of the space. 
Without this geometric prior, these clusters can be hard to identify, which highlights the importance of geometry in data science.
Moreover, the proposed method rests on a continuous (over-parametrized) relaxation to a discrete optimization problem, i.e., clustering.
The team applies their method to the clustering of neurons, but the application could extend into many fields.
Experiments are performed on simulated data, which they generated using \texttt{Geomstats}.

\paragraph{NeuroSeed} targets the field of genetics -- more specifically, smORFs. smORFs are small proteins that are used in cells across an enormous range of functions. The ``distance between small-proteins" is an important quantity because it will (i) bring us closer to classifying small proteins, and (ii) help us identify which genetic sequence each small protein comes from. Existing methods only analyze ``smORFs distances" in Euclidean space. The NeuroSEED submission analyzes smORF distances in hyperbolic space and proves that this approach is more efficient and yields more accurate results. They use \texttt{Geomstats'} \texttt{PoincareBall} (a hyperbolic manifold) to calculate hyperbolic distance with a novel unsupervised manifold embedding method. Their model uses a recurrent neural network to produce an encoder that has learned how to accurately estimate the Levenshtein Distance between two given sequences and embed them onto a given topology. Experiments are performed on the SmORFinder dataset \citep{Durrant2021}. 
Their results illustrate that hyperbolic spaces are powerful geometries for representing hierarchies in data.

\paragraph{Geomstats implementation} Both submissions are rather tailored to their applications, which makes them extremely powerful for their respective fields. It would be interesting to abstract them and generalize them for integration into \texttt{Geomstats}. For example, TEDDY could be structured to identify a wider range of topologies in datasets. NeuroSEED could help us create a general embedding encoder, whose goal is to respect a distance matrix given as input.

\section{Take-Aways for Geomstats}\label{sec:features}

\paragraph{Sampling} In contrast to other existing libraries coupling probability and geometry, such as \texttt{Geoopt} \citep{Kochurov2019Geoopt:Optim}, Geomstats does not have a sampling module. The contribution ``Brownian motion on manifolds" opens the door to creating such a module. In tandem with a sampling module, we could consider extending Geomstats' probability module that provides probability densities on manifolds.

\paragraph{Data sets} While \texttt{Geomstats} provides open-source data sets as resources for program testing, we saw that many participants used data sets from outside sources. This indicates that \texttt{Geomstats} could be improved by providing more sample data sets on manifolds. The submissions of the challenge pulled new open-source manifold-valued data sets that we can use to enrich the library's available data sets:
\begin{itemize}
    \item Synthetic data sets of neural signals,
    \item SmORF data sets (which can be embedded in hyperbolic manifolds),
    \item CS PhD data sets (which can be embedded in hyperbolic manifolds).
\end{itemize}

\paragraph{Data visualization with manifold embedding} Many of the submissions provided excellent examples for manifold embedding. For example, the Tree Embedding submission demonstrated how to embed a data set into a tree metric, and Rie-SNE demonstrated how to visualize high-dimensional data clustering by embedding the clusters into a two-dimensional image. Both serve as excellent resources for visualizing high-dimensional manifold data.

\paragraph{Manifold regression} WPGR provides a method that can perform regression on Riemannian manifolds for trajectories that are not geodesics. This is an extremely valuable tool, and the team has implemented this into \texttt{Geomstats}.

\paragraph{New metric implementation} The Sasaki team is currently working on implementing their \texttt{SasakiMetric} class in \texttt{Geomstats} as a subclass of the \texttt{RiemannianMetric} class.

\section{Final ranking}\label{sec:rank}

The Condorcet method was used to rank the submissions and decide on the winners. Each team whose submission respected the guidelines was given one vote in the decision process. The other judges were selected Geomstats maintainers and collaborators. 

Each team and each judge was asked to vote based on the evaluation criteria listed earlier in this document. Each team and each judge put in a vote for the three best submissions. Each of the three preferences had to be different: e.g. one could not select the same \texttt{Jupyter Notebook} for both first and second place. The votes remained secret. Only the four highest ranking submissions are published here:

\begin{enumerate}
    \setlength\itemsep{-0.2em}
    \item Hyperbolic Embedding via Tree Learning,
    \item Wrapped Gaussian Process Regression on Riemannian Manifolds,
    \item Riemannian Stochastic Neighbor Embedding,
    \item Sasaki Metric and Applications in Geodesic Analysis.
\end{enumerate}

Regardless of this final ranking, we would like to stress that all the submissions were of very high quality. We warmly congratulate all the participants.

\subsubsection*{Author Contributions}

Nina Miolane led the organization of the challenge. Nina Miolane, Saiteja Utpala, and Shubham Talbar were responsible for the GitHub repository. Adele Myers analyzed and summarized the results of the challenge. Adele Myers, Saiteja Utpala, Shubham Talbar, Sophia Sanborn, Christian Shewmake, Claire Donnat, Johan Mathe, and Umberto Lupo were the external reviewers in the evaluation process. The remaining authors of this white paper were the participants of the challenge.

\subsubsection*{Acknowledgments}

The authors would like the thank the organizers of the ICLR 2022 workshop ``Geometrical and Topological Representation Learning" for their valuable support in the organization of the challenge and specifically Bastian Rieck for his availability and help.

\section{Conclusion}

This white paper presented the motivations behind the organization of the ``Computational Geometric and Topological Challenge" at the ICLR 2022 workshop ``Geometric and Topological Representation Learning" and summarized the findings from the participants' submissions.

The submissions implemented methods for: data regression on manifolds (WGPR), manifold visualization through manifold embedding (Rie-SNE and Tree Embedding), clustering through manifold embedding (NeuroSEED), clustering through an unsupervised topology learning algorithm (TEDDY), and sampling from Brownian motion. Another submission implemented a new \texttt{SasakiMetric} class in Geomstats (Sasaki) and showcased learning algorithms on it. 

\bibliography{iclr2022_workshop}
\bibliographystyle{iclr2022_workshop}
\end{document}

%% file: math_commands.tex
\usepackage{amsmath,amsfonts,bm}

\def\eqref#1{equation~\ref{#1}}

\def\1{\bm{1}}

\DeclareMathAlphabet{\mathsfit}{\encodingdefault}{\sfdefault}{m}{sl}
\SetMathAlphabet{\mathsfit}{bold}{\encodingdefault}{\sfdefault}{bx}{n}

